# Implementation of Near-Surface Disjoining Pressure Effect in Continuum Simulations

Sajag Poudel, An Zou, Shalabh C. Maroo*

Department of Mechanical & Aerospace Engineering, Syracuse University, NY 13244

*scmaroo@syr.edu

**Abstract:** An expression of disjoining pressure in a water film, as a function of distance from the surface, is developed from prior experimental findings. The expression is implemented in commercial computational fluid dynamics solver and disjoining pressure effect on water wicking in nanochannels of height varying from 59 nm to 1 micron is simulated. The simulation results are in excellent agreement with experimental data, thus demonstrating and validating that near-surface molecular interactions can be integrated in continuum numerical simulations through the disjoining pressure term. Such an implementation can be used to design and advance nanofluidics-based engineering systems.

**Main Text:**

The absolute pressure in a nanoscale thin liquid film is significantly reduced from bulk due to solid-liquid interatomic interactions; such a reduction is characterized by the well-established disjoining pressure theory[1]. The effect of disjoining pressure on liquid film pressure is mathematically captured through the modified Young-Laplace equation.[2] Disjoining pressure occurs in a wide range of natural and engineering processes where thin liquid films are ubiquitous, such as in heat-transfer,[2-13] transpiration,[14-16] droplet-spreading,[5,6,17] and those involving bubbles,[2,18-20] etc. The quantitative contribution of interatomic interaction to disjoining pressure is dependent on the atomic composition and distance from the surface. It is mainly attributed to van der Waal's force if either solid or liquid atoms are non-polar; long-range electrostatic forces dominate when both solid and liquid atoms are polar. In this work, we focus on the latter scenario comprising water – silicon dioxide combination where electrostatic forces can affect molecular motion of water up to tens to hundreds of nanometers from the surface.[21-23]

In current literature, theoretical estimation of disjoining pressure of polar molecular combinations, i.e. water – silicon dioxide, using DLVO theory requires approximation of surface potential;[24] while discrete numerical simulations, such as molecular dynamics, are currently limited by the computational ability to simulate large domains. In our recent work,[25] the disjoining pressure of water was quantitatively characterized by conducting wicking experiments in 1-D silicon dioxide nanochannels. Disjoining pressure was found to dominate the driving mechanism for the liquid wicking in low height nanochannels ($h$ < 100 nm). The average value of disjoining pressure ($\widehat{P_d}$) in the water film was deduced as an exponential function of the film thickness ($\delta$) (i.e. half of the channel height);[25] however, due to the challenges of measuring local pressure at locations away from the surface at the nanometer scale, the knowledge of the disjoining pressure distribution in water film as a function of the distance ($y$) from the surface, $P_d(y)$, is still lacking[1,26] along with its integration in continuum simulations, and is the focus of this work. We first develop an expression to estimate $P_d(y)$, following which we implement the disjoining pressure effect in a commercial computational fluid dynamics (CFD) solver by supplementing an additional source term in the momentum equation. The implementation is then utilized to simulate the wicking process of water in nanochannels connected to a reservoir (Fig. 1a). Our work integrates and validates a molecular level phenomenon (disjoining pressure) into continuum simulations, and can be further used to study nanoscale effects as well as design engineering systems involving nanoscale liquid flows.

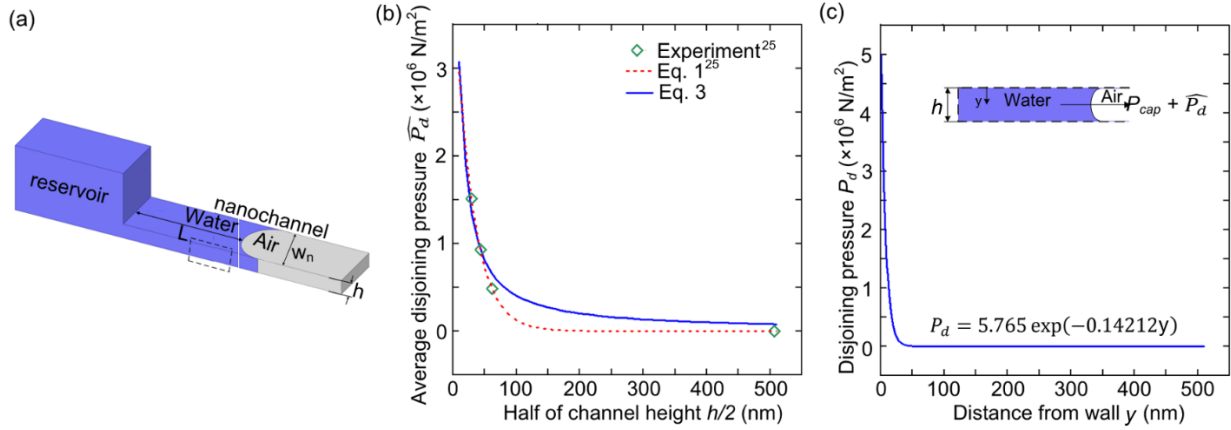

*Figure 1. Wicking in nanochannels. (a) Schematic of water wicking in a nanochannel connected to a reservoir. (b) Variation of average disjoining pressure existing in a nanochannel with channel half-height. (c) Variation of disjoining pressure with distance from the wall.*

Equation 1 represents the average value of disjoining pressure of water in nanochannel ($\widehat{P_d}$) as a function of the liquid film thickness (=$h/2$ which is half the nanochannel height) obtained from curve fit of experimental data in our recent work[25] (Fig. 1b). In order to develop an expression of $P_d$ as a function of the distance from the surface ($y$), we adopt a similar exponential function as shown in Eq. 2, where $A$ and $B$ are constants to be determined. By integrating Eq. 2 with respect to $y$ as shown in Eq. 3, the average disjoining pressure of water in nanochannel can be obtained, and then compared against Eq. 1 to estimate the constants as explained next.

$$\widehat{P_d} = 4.25 \exp\left(-0.035 \frac{h}{2}\right) \qquad \text{Equation 1}$$

$$P_d = A \exp(-By) \qquad \text{Equation 2}$$

$$\widehat{P_d} = \frac{1}{h}\int_0^h P_d(y)\, dy = \frac{1}{h}\int_0^h A\exp(-By)\, dy = \frac{A[1-\exp(-Bh)]}{hB} \qquad \text{Equation 3}$$

Constants $A$ and $B$ were determined by an iterative process. First, random values were assigned in Eq. 3, whose results were compared to that from Eq. 1. The values were converged in each iteration by minimizing the error between $\widehat{P_d}$ computed from Eq. 1 and Eq. 3. The convergence is obtained by estimating the coefficient of determination $R^2$ between the solutions of Eqs. 1 and 3 for the corresponding $h$. The best fits of $A$ and $B$ are selected for the highest and the most stable value of $R^2$ (i.e. the value of $R^2$ which would be independent of the number of data-points considered). Eventually, constants $A$ and $B$ were finalized as 5.765 and 0.14212 respectively (see Eq. 4), with $R^2$ = 91.3% resolved to the precision of $\Delta y$ = 1 nm with ~500 data points.

$$P_d = 5.765 \exp(-0.14212y) \qquad \text{Equation 4}$$

Substituting these values of A and B, the averaged disjoining pressure ($\widehat{P_d}$) from Eq. 3 is in good agreement with both experimental data and fitting curve (Eq. 1) prediction (Fig. 1b). The corresponding disjoining pressure distribution $P_d$ in water film is plotted in Fig. 1c. Further information on the sensitivity of $R^2$ with the number of data points is available in supplemental material section S1.

Next, to simulate the wicking process in nanochannels of geometry consistent with experiments[25] (Table 1), disjoining pressure effect is implemented in continuum simulation using a laminar multiphase model with the volume of fluids method. A commercial CFD software ANSYS Fluent is utilized to solve the governing equation of fluid flow and a user defined function (udf) is linked to the momentum equation to specify the source term based on the disjoining pressure function.[27,28] As shown in Fig. 2-a1, a nanochannel of height ($h$), length ($L_n$), and width ($w_n$) is connected to the reservoir of height ($h_m$), length ($L_m$), and width ($w_m$). $L_m$ and $w_m$ were fixed as 150 µm and 40 µm respectively; while $h_m$ changed with nanochannel height ($h$) with a fixed ratio ($h_m/h$) of ~30. Four channel heights ($h$) of 59 nm, 87 nm, 124 nm, and 1015 nm were simulated with fixed channel length $L_n$ of 150 µm, and width $w_n$ of 10 µm (Table 1). Based on the advantage of symmetry, only one-fourth portion of the physical domain is considered for the CFD simulation (Fig. 2-a2). The faces AB and EF towards the reservoir and the nanochannel respectively are open with the pressure outlet (1 atmosphere) boundary condition. The face ADF is symmetric and no-slip condition is set at all walls of nanochannel and reservoir. Surface tension is evoked using a continuous shear force model[29] (liquid-air surface tension $\gamma$ = 0.0072 N/m) together with wall adhesion and specified contact angles. The contact angles on the side and top walls of each case of nanochannel height are adopted from our prior experimental work[25] and are listed in Table 1 (*see Supplemental Material Section S2*). Figure 2-a3 illustrates a magnified view of the XY plane of the domain with the plane of symmetry (green line) and Fig. 2-a4 demonstrates mesh refinement near the top wall of the nanochannel.

*Table 1. Nanochannels geometry and corresponding contact angles at top and side walls* [25]

| Case | Reservoir (Microchannel Geometry) | | Nanochannel Geometry | | Contact Angle on Nanochannel walls | |
|---|---|---|---|---|---|---|
| | Height ($h_m$) | | Channel height ($h$) | Channel Length ($L_n$) = 150 µm, Channel width ($w_n$) = 10 µm | Top Wall | Side Wall |
| 1 | 1.7 µm | Length ($L_m$) = 150 µm, Width ($w_m$) = 40 µm | 59 nm | | 90° | 29.4° |
| 2 | 2.5 µm | | 87 nm | | 90° | 27.3° |
| 3 | 3.5 µm | | 124 nm | | 64.8° | 40.6° |
| 4 | 30 µm | | 1015 nm | | 39.6° | 39.6° |

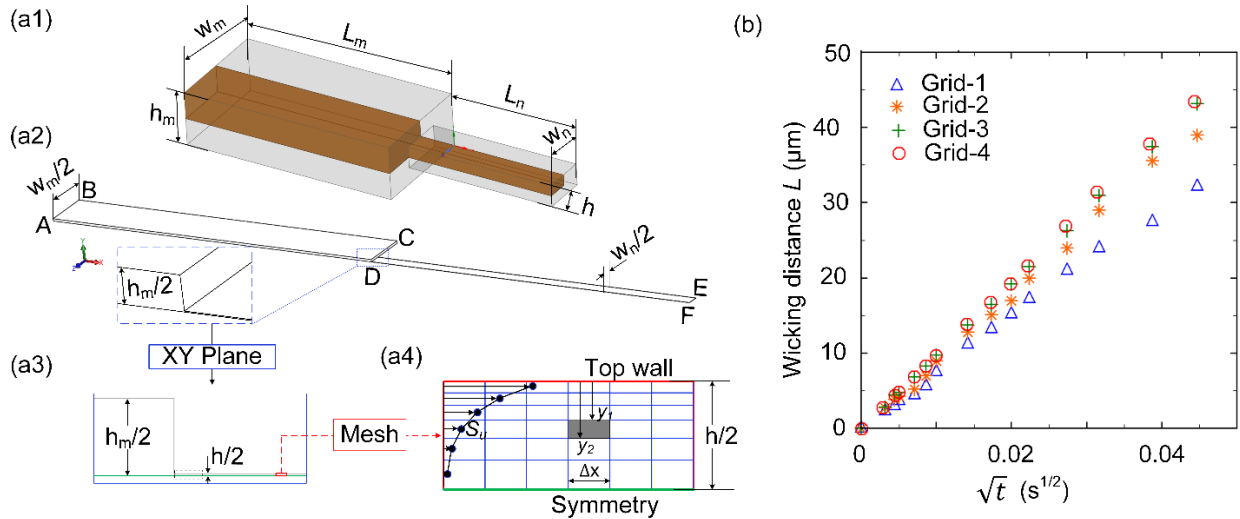

*Figure 2. CFD Simulation of water wicking in a nanochannel. (a1) Sketch of a nanochannel connected to a reservoir showing symmetry. (a2) Computational domain utilized in the present study. (a3) XY plane view of the domain showing the plane of symmetry. (a4) Mesh refinement near the wall of nanochannel together with the specified source term profile along the height. (b) Comparison of wicking distance evolution with time for simulation in nanochannel h = 59 nm by using different levels of mesh refinement.*

The effect of the disjoining pressure (Eq. 4) is included by adding a source term $S_u$ in the liquid phase, which is added to the discretized form of the X-momentum equation through a udf as follows:

$$S_u = \int_{y_1}^{y_2} \frac{P_d}{\Delta x} \qquad \text{Equation 5}$$

where, $y_1$ and $y_2$ are limits of the finite volume cell in Y-axis and $\Delta x$ is the cell width along X-axis (see Fig. 2-a4). The stated source term in the corresponding momentum equation is estimated at the center of each finite volume cell and acts as a supplementary driving force for liquid propagation. As shown in Fig. 2-a4, the supplied source term wizll clearly be dependent on grid-spacing. A mesh with non-uniform grid spacing along the channel height is found to better include the disjoining pressure effect on wicking as compared to one with uniform gird. Thus, a mesh with refined grid spacing near the wall is employed in all cases of simulation considered (see Supplemental Material Section S3).

A grid-independent study was first performed by varying the number of finite volume cells along the channel height, using different meshes. For the 59 nm channel, four different meshes with non-uniform grid spacing were utilized, with 4, 5, 6, and 7 finite volume cells along the channel half-height respectively (denoted as Grid-1, Grid-2, Grid-3, and Grid-4). During the wicking process, the location of the meniscus at a time instant was identified by extracting the volume fraction of liquid phase ($v_l$), with $v_l \sim 0.5$ used to identify the meniscus. The distance between the entrance of the channel and the meniscus was defined as wicking distance *L* (Fig. 1a). The evolution of wicking distance *L* with time *t* served as the criteria to determine the simulation reliability. As seen in Fig. 2b, the result of Grid-3 (6 finite volume cells) is consistent with the result of Grid-4 (7 finite volume cells), indicating that Grid-3 is an optimum one. Moreover, an analysis of Richardson's error estimation[30] is employed to obtain the relative error in simulation with different grids, which also revealed that the error in simulation results based on Grid-3 is minimum with extensive grid convergence (see Supplemental Material S3). Thus, meshing based on Grid-3 is used and wicking simulations are performed for different channel heights.

Figure 3 shows the evolution of wicking distance *L* with time from CFD simulations for all four channel heights; inset in the figures show the spatial variation of the additional source term $S_u$. By including the effects of both $P_{cap}$ and $P_d$, the simulation data are in good agreement with experimental data.[25] Thus, the expression of disjoining pressure $P_d(y)$, as a function of distance from the surface $y$ (Eq. 2), for water – silicon dioxide combination, as well as the implementation of disjoining pressure effect to CFD solver is demonstrated to be accurate as they adequately reproduce the experimental outcome. The findings from the present work can be used to further study phenomenon where near-surface effects on liquid behavior are non-trivial.

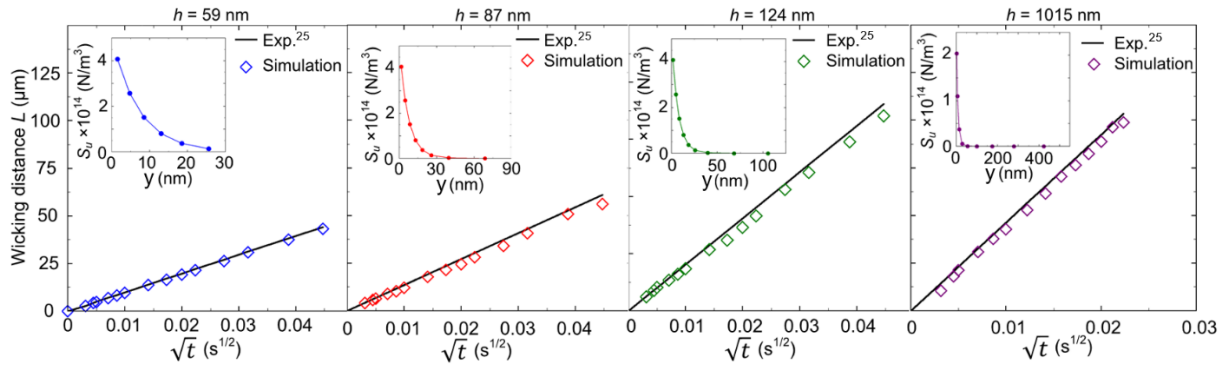

*Figure 3. Evolution of wicking distance with time for nanochannels of varying height.* The inset in each plot shows the spatial variation of $S_u$, where the number of data points (solid circle) in each denotes the corresponding number of finite volume cells along $h/2$, and the horizontal axis indicates the distance of the finite volume cell center from the top wall of the channel.

Figure 4 shows the variation of pressure along the direction of wicking in 59 nm and 1015 nm channels. During this process, there exists a pressure gradient along the direction of wicking with a peak negative pressure ($P_{np}$) at the meniscus,[6,8,31] and is estimated from the corresponding contour plots (insets of Fig. 4) of liquid volume fraction inside the channel. The capillary pressure is obtained by $P_{cap} = 2\sigma \left( \frac{\cos\theta_{side}}{h} + \frac{\cos\theta_{top}}{w} \right)$, with contact angle values listed in Table 1; while the value of average disjoining pressure is obtained from Eq. 2. Table 2 lists all values of $P_{cap}$, $\widehat{P_d}$ and $P_{np}$ for all channel heights. For small channel heights (<100 nm), $P_d$ dominates and thus the relation $P_d \approx P_a - P_{np}$ holds; while in 1015 nm channel, $P_d$ is negligible and thus the relation $P_{cap} \approx P_a - P_{np}$ applies. From this investigation of liquid film pressure, the relative contribution of $P_{cap}$ and $P_d$ on the overall driving mechanism is found to widely vary for the two extreme heights of nanochannels considered in the present study.

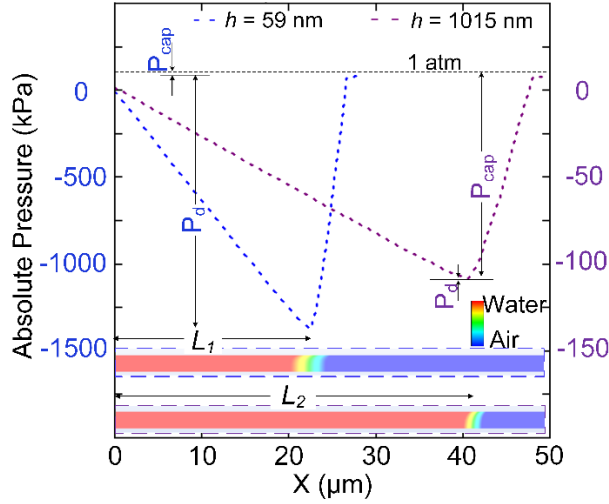

*Figure 4. Absolute pressure along the nanochannel length indicating the peak negative pressure at the meniscus. The inset (corresponding to the dashed-box of Fig. 2(a3)) shows the liquid phase volume fraction contour plot of the nanochannel.*

*Table 2. The values of different pressures associated with each case of nanochannel height.*

| Channel height (h) | $P_{cap}$ | $\widehat{P_d}$ | $P_{np}$ |
|---|---|---|---|
| 59.6 nm | 1.25×10⁴ | 1.39×10⁶ | -1.40×10⁶ |
| 87 nm | 1.28×10⁴ | 9.30×10⁵ | -8.78×10⁵ |
| 124 nm | 4.46×10⁵ | 6.54×10⁵ | -8.23×10⁵ |
| 1015 nm | 1.09×10⁵ | 7.91×10⁴ | -1.05×10⁴ |

In summary, we utilize the experimental findings of average disjoining pressure of water in nanochannels and derive an expression for disjoining pressure distribution in a water film as a function of distance from the surface. The developed disjoining pressure expression is implemented in computational fluid dynamics solver to perform simulations of wicking in nanochannels to include near-surface molecular effects. The simulation results are in excellent agreement with the experimental data, demonstrating the successful implementation and validation of disjoining pressure in continuum simulations. The disjoining pressure in the liquid film was found to be the primary driving mechanism for the flow in nanochannels for height < 100 nm. The present work lays a foundation for future continuum studies of physical phenomenon where nanoscale liquid films are prominent.

**Acknowledgment:** This material is based upon work supported by, or in part by, the Office of Naval Research under contract/grant no. N000141812357. The numerical simulations were performed in the UberCloud virtual machines, an official cloud hosting partner of ANSYS.

**Data Availability Statement:** The data that support the findings of this study are available from the corresponding author upon reasonable request.